\documentclass[preprint2]{aastex6}
\usepackage{natbib}
\usepackage{graphicx}
\usepackage{wasysym}
\usepackage{txfonts}
\def\fH2{\mbox {f$_{{\rm H}_2}$}}

\def\EBV{\mbox{E$_{\rm B-V}$}}

\def\AV{\mbox{A$_{\rm V}$}}

\def\nH2{\mbox{${\rm n}(\HH$)}}
\def\enH2{\mbox{$n_{(\HH$)}}}

\def\pccc{~{\rm cm}^{-3}} 
 
\def\pcc {\mbox{${~{\rm cm}^{-2}}$}}

\def\Tsub#1 {\mbox{${\rm T}_{\rm #1}$}}
\def\TK  {\Tsub K }

 \def\arcmin{\mbox{$^{\prime}$}}

\def\degr{$^{\rm o}$}
\def\p{\mbox{$^+$}}
\def\m{\mbox{$^-$}}

\def\cch{\mbox{C$_2$H}}
\def\cfh{\mbox{C$_4$H}} 
\def\cfhm{\mbox{C$_4$H$^-$}}

\def\h13cop{\mbox{{H$^{13}$CO\p}}}

\def\c3h2{\mbox{C$_3$H$_2$}}
\def\linearC3H{\mbox{$l$-C$_3$H}}
\def\linear{\mbox{$l$-C$_3$H$_2$}}
\def\cyclic{\mbox{$c$-C$_3$H$_2$}}
\def\cyclicC3H{\mbox{$c$-C$_3$H}}
 
 \def\R0{R$_0$}
\def\G0{\mbox{G$_0$}}

\def\ddeg{{}^\circ\kern-.1em}

\def\kms{\mbox{km\,s$^{-1}$}}
\def\ps{\mbox{s$^{-1}$}}
\def\bll{BL Lac}

\def\E#1 {$10^{#1}$}
\def\E#1 {E{#1}}
\def\P#1,{$\nH2\TK~=~#1\times~10^4\pccc$~K}
\def\ec#1,#2,#3,{#1\,(#2)\E{#3}}

\def\HC3N{\mbox{HC$_3$N}}
\def\methCN{\mbox{CH$_3$CN}}
\def\methNC{\mbox{CH$_3$NC}}
\def\methCCH{\mbox{CH$_3$C$_2$H}}
\def\H3{\mbox{H$_3$}}

\def\RH2{\mbox{R$_{\rm G}$}}
\def\g13{\mbox{g$_{13}$}}

\def\kHeH2{\mbox{$k_{ He-\HH}$}}

\def\tim#1,#2{\mbox{{$#1\times10^{#2}$}}}

\newcommand{\emm}[1]{\ensuremath{#1}}   
\newcommand{\emr}[1]{\emm{\mathrm{#1}}} 

\newcommand{\hcop}{\emr{HCO^+}} 
 
\newcommand{\HH}{\emr{H_2}}

\slugcomment{generated \today}

\shorttitle{Small hydrocarbons \& \methCN\ in local  diffuse molecular gas}
\shortauthors{Liszt, Gerin, Beasley \& Pety}

\begin{document}


\title{Chemical complexity in local diffuse and translucent clouds: 
ubiquitous linear-C$_3$H and CH$_3$CN, a detection of HC$_3$N and an 
upper limit on the abundance of C\HH CN}

\thanks{Based on observations obtained with the NRAO Jansky Very Large Array (VLA)}


\author{Harvey Liszt} 
\affil{National Radio Astronomy Observatory \\
        520 Edgemont Road, Charlottesville, VA 22903-2475 \\
       hliszt@nrao.edu}
\and
\author{Maryvonne Gerin}
\affil{LERMA, Observatoire de Paris, PSL Research University, \\
 CNRS, Sorbonne Universit\'e, UPMC Universit\'e Paris 06, \\
 Ecole Normale Superieure, 75005, Paris, France \\
 gerin@lra.ens.fr}
\and

\author{Anthony Beasley}
\affil{National Radio Astronomy Observatory \\
        520 Edgemont Road, Charlottesville, VA 22903-2475 \\
  tbeasley@nrao.edu}
\and
\author{Jerome Pety}
\affil{ Institut de Radioastronomie Millim\'etrique,
        300 Rue de la Piscine, F-38406 Saint Martin d'H\`eres, France \\ 
     CNRS, Sorbonne Universit\'e, UPMC Universit\'e Paris 06, \\
     Ecole Normale Superieure, 75005, Paris, France \\
  pety@iram.fr
}




\begin{abstract}

We present Jansky Very Large Array observations of 20 - 37 GHz absorption
lines from nearby Galactic diffuse molecular gas seen against four 
cosmologically-distant compact radio continuum sources.  The main new 
observational results 
are that \linearC3H\ and \methCN\ are ubiqitous in the local diffuse molecular 
interstellar medium at \AV\ $\la 1$ while HC$_3$N was seen only toward B0415 
at \AV\ $>$ 4 mag.  The linear/cyclic ratio is much larger in C$_3$H
than in C$_3$\HH\ and the ratio \methCN/HCN is enhanced compared to TMC-1,
although not as much as toward the Horsehead Nebula.  More consequentially, this 
work completes a long-term
program assessing the abundances of small hydrocarbons (CH, \cch, linear and 
cyclic C$_3$H and C$_3$\HH,  and \cfh\ and \cfhm) and the CN-bearing species
(CN, HCN, HNC, HC$_3$N, HC$_5$N and CH$_3$CN):  their systematics in diffuse 
molecular gas are presented in detail here.  We also observed but did not strongly 
constrain the abundances of a few oxygen-bearing species, most prominently 
HNCO. We set limits on the column density of C\HH CN, such that the 
anion C\HH CN\m\ is only viable as a carrier of diffuse interstellar bands if the 
N(C\HH CN)/N(C\HH CN\m) abundance ratio is much smaller in 
this species than in any others for which the anion has been observed.
We argue that complex organic molecules are not present  in clouds meeting
a reasonable definition of diffuse molecular gas, ie \AV\ $\la 1$ mag.
\end{abstract}



\section{Introduction}

The molecular inventory of diffuse interstellar gas is interesting
because the unexpectedly high abundances of trace species imply the 
presence of underlying physical processes that might otherwise remain 
hidden \citep{GodFal+14}.  But knowledge of the molecular complement of 
diffuse molecular gas can be used to advantage  even when the underlying physical 
processes and observed abundances are only very imperfectly understood:  

\begin{itemize}

\item Chemistry provides reliable \HH-tracers with
well-determined relative abundances from optical astronomy such as 
OH (X(OH) = N(OH)/N(\HH) $= 10^{-7}$ \citep{WesGal+09,WesGal+10})
and CH (X(CH) $= 3.5\times 10^{-8}$ \citep{SheRog+08}) as well as
\hcop\ that is observed in absorption at 89 GHz with an abundance 
X(\hcop) $= 3\times 10^{-9}$ that can be fixed with respect to both 
CH and OH \citep{LisLuc96,LisPet+10,LisGer16}.  

\item The empirically-determined relative abundance of \hcop\ suffices 
to explain observations of widely-observed CO in diffuse molecular gas 
\citep{LisPet+10} as  the product of  recombination of \hcop\ with ambient 
electrons \citep{GlaLan75,Lis07CO,VisVan+09,Lis17CO} followed by exchange of 
carbon isotopes \citep{WatAni+76,Lis17CO}.

\item Tallying the inventory of identifiable molecular species sets 
broad guidelines for attributing practicable carriers of diffuse 
interstellar bands (DIBs) \citep{LisSon+12,LisLuc+14}. We 
recently showed that \linear\ is not sufficiently abundant 
to serve as the carrier of the DIBs at 4881\AA\ and 
5450\AA\ with which it was tentatively identified on the 
basis of coincidences in laboratory spectra \citep{MaiWal+11}. 
Constraining the abundance of another putative DIB-carrier, C\HH CN$^-$ 
\citep{CorSar07}, is one aspect of the present work.  Knowledge of the 
abundances of smaller molecules should help in understanding the 
abundances of broad groups of much larger species like polycyclic 
aromatic hydrocarbons (PAHs) that may not be individually identifiable.

\end{itemize}

The molecular inventory of diffuse molecular gas has recently been 
greatly enlarged using high spectral resolution heterodyne techniques.
The HiFi instrument on HERSCHEL observed an extensive inventory of 
carbon, oxygen and nitrogen hydrides and hydride ions 
in the sub-mm and THz domains \citep{GerNeu+16}, including species 
long known in optical absorption (CH, CH\p) but also such species as 
hydrofluoric acid (HF) and the argonium ion ArH\p .
CF\p, $c-$C$_3$H and HCO were detected in local gas at the IRAM 
30m telescope \citep{LisPet+14} and CF\p\ was subsequently detected 
in diffuse molecular gas across the galactic disk \citep{LisGuz+15} 
using NOEMA.

In this work we were motivated to extend the molecular inventory
and explore the limits of chemical complexity in diffuse molecular 
gas, given the recently-developed spectroscopic capabiliies of the 
enhanced Jansky Very Large Array (VLA).  We used the VLA to search
at frequencies 20 - 37 GHz for polyatomic molecules whose transitions 
are most favorably observed in the cm-wave band.  We study three 
chemical families:

\begin{itemize}

\item Hydrocarbons.  \linear\ and \cyclic\ are the heaviest molecules 
known in local diffuse molecular gas but they and c-C$_3$H are as 
ubiquitous as the lighter hydrocarbons CH and \cch : by contrast, 
\cfh\ has not been detected \citep{LisSon+12,LisLuc+14}.  Here we 
demonstrate the ubiquity and
high abundance of \linearC3H\ and compare the abundances of the linear 
and cyclic versions of C$_3$H and C$_3$\HH .  \cite{LoiAgu+17}
have recently shown that the relative abundance of the linear
and cyclic versions of these molecules represents a competition between
formation and isomerization by interaction with atomic hydrogen.
The abundance of neutral atomic hydrogen is much higher in diffuse 
molecular gas, presenting an interesting test of the chemistry.  
\cite{LoiAgu+17}
stress the role of C$_3$, which is uniquely observable in diffuse
molecular gas. Here we show that C$_2$, also uniquely observable
in diffuse molecular gas, is by a slight margin over CH and \cch\ 
the most abundant carbon-bearing molecule after CO: this would 
not have been possible without a comprehensive survey.

\item CN-bearing molecules.  The relative abundances of CN, HCN and HNC
are very nearly constant in diffuse molecular gas \citep{LisLuc01}
but larger CN-bearing species have yet to be detected  \citep{LisPet+08}.
Here we show that \methCN\ is ubiquitous at \AV\ = 1 mag, which is 
quite a surprise given that recent models of the formation of \methCN\ at such
moderate extinction predicted an abundance of \methCN\ that is some five 
orders of magnitude below the observed levels \citep{MajDas+14}.  We also 
detect HC$_3$N toward B0415+379 (3C111) at \EBV\ = 1.6 mag but not toward 
B2200+420 (\bll) at \EBV = 0.33 mag (\EBV\ $\simeq$ \AV/3.1).

\item Oxygen-bearing molecules.  Previously-detected, lighter oxygen-bearing 
species include OH, \hcop, HOC\p, HCO and CO observed at radio frequencies, 
and \HH O and the many oxygen hydrides and hydride ions observed by HiFi 
\citep{GerNeu+16}. \HH CO, usually thought to form 
on dust, is known to be ubiquitous in diffuse molecular gas 
\citep{Nas90,MarMoo+93,LisLuc+06} although CH$_3$OH, which must form 
on grains, is not detected \citep{LisPet+08}.  Here we set limits on an 
eclectic group of heavier oxygen-bearing species HNCO, HCOOH (formic acid) 
and \HH COH\p.  The systematics of the oxygen-bearing species are not 
discussed here, owing to the paucity of significant new results.

\end{itemize}

The plan of this work is as follows.  In Section 2 we describe the new 
observations discussed here and the manner of the presentation of the
results.  In Sections 3 - 5 we separately discuss the hydrocarbon, 
CN-bearing and oxygen-bearing species including results for CH, CN,
C$_2$ and C$_3$ that are observed in optical/UV absorption along 
sightlines having comparable \EBV\ and CH to those observed here.  
Section 6 discusses the
viability of C\HH CN\m\ as a DIB carrier, Section 7 compares our
results with those of \cite{ThiBel+17} for diffuse clouds observed
toward Sgr B2 in the Galactic center and disk and Section 8 is a 
summary.

\section{Observations, conventions and conversion from optical depth to column density}

\subsection{Observing and data reduction}

The new observations reported here were taken at the National Radio
Observatory's Very Large Array (VLA) on 17 June and 6 July 2013 under 
proposal 13A-097 while in the C-configuration having angular resolution
25-45 milliarcsec.  The data were taken in four 
scheduling blocks (SB) of 2 hour duration, observing absorption against the 
continuum targets listed in Table 1 in two orthogonal polarizations.   The 
observing was done with 8 spectral windows having 512 channels of resolution and 
separation 78 kHz placed opportunistically within the range 20.1 - 22.5 GHz in 
June (corresponding to velocity resolution 0.104 - 0.115 \kms) and 512 channels of 
resolution and separation 156 kHz within the range 32.7 - 37.3 GHz in July,
corresponding to velocity resolution 0.126 - 0.143 \kms .  Spectroscopic
properties of the newly-observed spectral lines discussed here are summarized 
in Table 2. 

\begin{figure*}
\includegraphics[height=7.4cm]{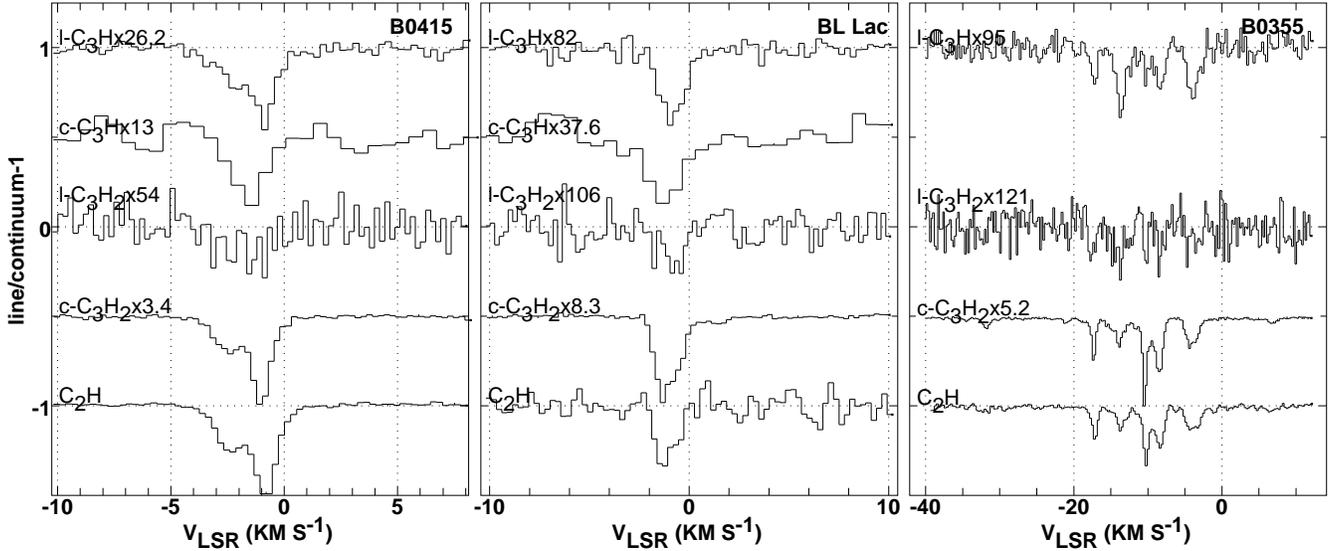}
  \caption{ Hydrocarbon line profiles toward B0415+379 ,B2200+420 and B0355+508 ,
vertically displaced and scaled as indicated. The \linearC3H\ profiles at the top of each
panel are new from this work. c-C$_3$H was not observed toward B0355.}
\end{figure*}

As in our earlier project discussed in \cite{LisSon+12}, two continuum targets and 
a bandpass calibrator (3C84) were covered in each SB.
Considerable time was devoted to reference pointing on each continuum source 
before it was observed.  No absolute amplitude calibration was performed but 
the fluxes relative to that of the bandbass calibrator 3C 84 
(S$_\nu \approx 10-16$ Jy) are given in Table 1. In each SB the bandpass calibrator 
was observed for approximately 20 minutes.  The sources were observed for approximately 
40 minutes during any one SB execution.  

The data were calibrated using very standard techniques in CASA, largely repeating the
procedures described in \cite{LisSon+12}: Overall the scheme resembles that given
in online CASA tutorials for spectral line sources such as TW Hydra with
the notable exception that each absorption target, being a phase calibrator,
serves as its own phase calibrator.  The bandpass 
calibrator observations were phase-calibrated within each scan sub-interval, 
followed by construction of an average bandpass solution.  This was applied on 
the fly to complex gain-cal solutions for each continuum target at the sub-scan 
level, followed by scan-length gain calibration solutions to be  applied to 
each target individually.  Once the data were passband- and phase-calibrated in 
this way they were also fully reduced given the point-like nature 
of the  background targets.  For each polarization and baseband, 
spectra were extracted as vector phase averages over all visibilities,
without gridding, mapping or, indeed, more than very minimal manual flagging
of bad datapoints.  The spectra were 
produced in CASA's plotms visualizer and exported to 
drawspec singledish software \citep{Lis96}
where spectra in the two polarizations 
were co-added and very small linear baselines amounting typically to 
0.01\% of the continuum were removed from each of the basebands.  

 All velocities 
discussed here are relative to the kinematic definition of the Local 
Standard of Rest that is in universal use at radio telescopes.

\subsection{Conversion from integrated optical depth to column density}

Equivalent widths (integrated optical depths) are given in Table 3.
For \linearC3H the entries are the average of the two lines observed.
For \methCN\ the entries for K=0 are the sum of the optical depths of the
three K=0 transitions listed in Table 1.  The K=0 and K=1 transitions are
easily distinguished toward B0415 and B2200 (Figure 4) but not toward B0355, given
the complex kinematic structure and modest signal/noise.  For B0355 the only
quantity given in Table 3 is the integrated optical depth for all
kinematic components summed over both K-ladders and the total column density
was determined by scaling with respect to the analogous quantity derived
toward the sources B2200 and B0415 where the K-ladder structure was resolved..
 
Default factors needed to convert the observed integrated optical depths
(Table 3) to column density (Tables 4-6) are given in the next-to-last column 
of Table 2: these were computed by 
assuming rotational excitation in equilibrium with the cosmic microwave 
background. This is an excellent approximation for strongly-polar diatomics 
(ie, not CO) and smaller  polyatomics having low-J transitions in the 
mm-wave regime where emission is demonstrably weak, typically a few 
hundredths of a Kelvin for even optically thick lines 
\citep{LucLis96,LisPet16}. However, for lower-lying transitions of 
heavier species observed at cm-wavelengths as in this work, 
collisional excitation  more efficiently redistributes the 
rotational population out of the lowest states, increasing the numerical 
factors that should be used to convert observed optical depths to column 
density. 

An upward correction factor due to rotational excitation is tabulated
separately as a range in the right-most column of Table 2, corresponding
to results for the  density range n(\HH) $= 0 - 400 \pccc$ that is used 
in Appendix A.  The maximum correction is often below 2 but can be larger 
when the lowest-lying transition was observed.  We have kept the default
equivalent width - column density conversion separate from application
of the excitation correction, in part because all of the new observations 
are unlikely to be characterized by the same density, but 
the correction should be kept in mind during the discussion and it 
is noted explicitly in the text as required.  Throughout, we have 
avoided drawing conclusions that seemed unwarranted in the face 
of this uncertainty.

\subsection{Presentation of results: Figures and tables}

\begin{figure}
\includegraphics[height=7.7cm]{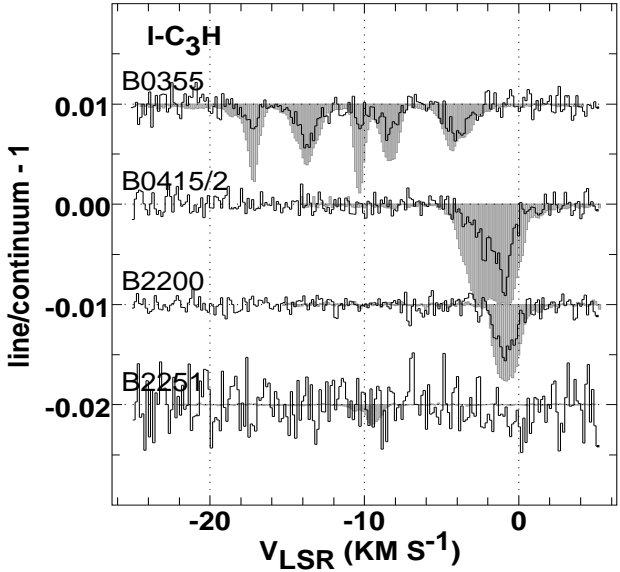}
  \caption{ Line profiles of \linearC3H\ for all sources are shown as histrograms 
compared with profiles of \hcop\ shown shaded in light grey and scaled downward by 
a factor 100. For B0415  the \linearC3H\ profile has been scaled downward
by a factor 2. \hcop\ absorption toward B2251+158 is at -9.6 \kms.} 
\end{figure}

\begin{figure*}
\includegraphics[height=14.5cm]{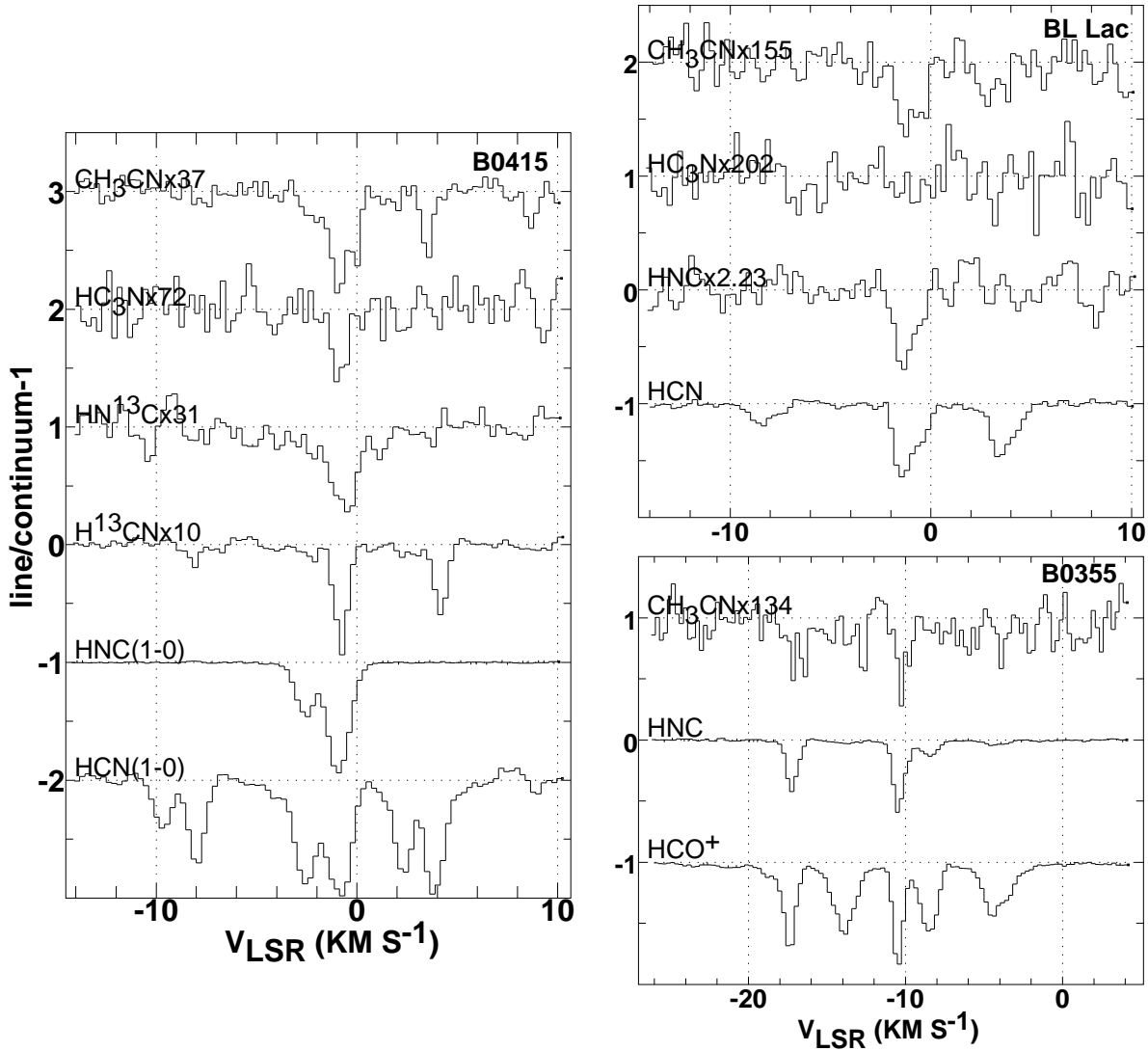}
  \caption{ Line profiles of nitrogen-bearing species toward B0415+379, 
B2200+420  and B0355+508, vertically displaced and 
scaled as indicated.  The \HC3N\ and \methCN\ profiles are new from this work.
The HCN spectrum is not included for B0355 owing to the complexity of the
kinematic structure.}  
\end{figure*}

Figure 1 shows results for the newly-detected species \linearC3H\  along 
with a complement of spectra of previously-observed hydrocarbons having two 
and three carbons: \cyclicC3H\ was not observed toward B0355+508 by 
\cite{LisPet+14}.  Figure 2 shows a source-by-source comparison of the 
newly-detected \linearC3H\ with spectra of \hcop, the species that shows 
the fullest extent of molecular absorption in our work. \linearC3H\ is 
clearly a very ubiquitous species in diffuse molecular gas but with 
substantial variation in the ratio of the strength of the observed 
transition to that of J=1-0 \hcop, as seen by comparing the individual
features seen toward B0355.  The expected variation
of the optical depth-column density conversion factor for \linearC3H\ is 
approximately 1 - 1.8 for densities in the number density range 
n(\HH) $= 0 - 400 \pccc$ (Table 2).   

Figure 3 compares spectra of the newly-detected CN-bearing species CH$_3$CN 
and HC$_3$N with those of previously-observed nitrogen-bearing species.  
The -17 \kms\ and -11 \kms\ components toward B0355+508 that are prominent 
in the CN-bearing species are just those that are weaker in \linearC3H\ in 
Figures 1 and 2.

Table 3 gives integrated optical depths for the newly-observed species listed
in Table 2 and Tables 4 - 6 give molecular column densities using the integrated 
optical depths in Table 3, calculated in the limiting case of no collisional
excitation above the cosmic microwave background.  For B0355+508 
the results are listed separately for the kinematic components that are known 
to exist toward this source in \hcop.   For the other sources, results are 
shown integrated across the velocity range of the \hcop\ profile. 

\subsection{Comparison with other millieu}

The results for diffuse clouds from our work are compared with abundances
for the same species determined in other environments in Tables 4 and 5
where detailed references are given, and in passing throughout the text. 
TMC-1 is the cyanopolyyne peak in the Taurus Dark Cloud, the well-known
Horsehead (HH) Nebula hosts a PDR and dense core that are distinguished 
in the tables.  B1b is a complex dark cloud core that has higher density 
and 3-10 times higher column density than TMC-1.  Abundances in the Orion 
Bar are as noted in the references in the tables.

\section{The abundances of small hydrocarbons}

We previously  showed that {\it c-}C$_3$H was ubiquitous in local
diffuse molecular gas \citep{LisPet+14} and the present work extends 
this statement to the linear variant \linearC3H. By contrast,
\cfh\ is not detected.  Based on the accumulated data shown in Table 4 and 
previous results for \cfh\ \citep{LisSon+12} we summarize the chemistry of 
small hydrocarbons as follows:

\begin{itemize}

\item \cch\ is generally the most abundant hydrocarbon. 
 N(\cch) $\ga$ N(CH) in diffuse molecular gas and dark cloud gas.

\item The fractional abundance of \cch\ is the same in diffuse molecular
gas ($4\pm 2 \times 10^{-8}$) and toward TMC-1 ($3-5 \times 10^{-8}$),
but much larger than toward B1b or the Horsehead environments
($0.3-1.0) \times 10^{-8}$.

\item Adding a third carbon beyond \cch\ to form C$_3$H produces a drop of 
about a factor 100 in column density in all environments.  The drop 
is larger in diffuse molecular gas (N(\cch)/N(C$_3$H) $\approx 200$) than 
in dark cloud gas  or the Horsehead (N(\cch)/N(C$_3$H) $\approx 30-70$) 
if {\it c-}C$_3$H is considered.  However, the drop is more nearly equal 
to 100 in all environments if the comparison is based on {\it l-}C$_3$H.

\item The cyclic/linear ratio N({\it c-}C$_3$H)/N({\it l-}C$_3$H) $\simeq 0.5 $ 
in diffuse molecular gas, comparable to what is observed in the circumstellar 
envelope around the evolved star  IRC+10216 (0.4, see \cite{AguCer+08B}), but very 
different from the values N({\it c-}C$_3$H)/N({\it l-}C$_3$H) $\simeq 3-10 $ 
in the other environments shown in Table 4.

\item The linear variant is much less abundant relative to cyclic
in C$_3$\HH\ than in C$_3$H.  The linear/cyclic ratio 
N(\linear)/N(\cyclic) $<< 1$ in diffuse molecular gas and the 
Horsehead environments, 1/40 - 1/15, and slightly larger, 1/7-1/6,
in dark cloud gas.

\item Abundance does not fall uniformly with complexity, \cch\ being
at least as abundant as CH, and \cyclic\ being more abundant than 
{\it c-}C$_3$H. N({\it l-}C$_3$H)/N({\it l-}C$_3$\HH) $\approx 1-3$ 
in all environments, and slightly larger in diffuse molecular gas than 
otherwise. N({\it c-}C$_3$H)/N({\it c-}C$_3$\HH) $\approx 1/10$ in 
diffuse molecular gas, only slightly less than in dark cloud gas (1/6-1/7). 
The Horsehead environents have  ratios nearer unity, 
N({\it c-}C$_3$H)/N({\it c-}C$_3$\HH) $\approx 1/2$

\item The ratios N(\linearC3H)/N(\cyclic) $ \approx 0.1 - 0.2$ observed 
here (Table 4) are quite comparable to those observed toward Sgr B2 
by \cite{CorMcG+17} in gas of indeterminate \EBV.

\item N(\cfh)/N(\cch) $<<1$ for diffuse molecular gas, smaller than
toward TMC-1 where N(\cfh)/N(\cch) $\approx$ 0.5 as we have summarized 
in Table 4 albeit with large uncertainty in N(\cfh) for TMC-1, see also 
\cite{LisSon+12}.  Our measurements of N(\cfh) are insufficiently 
sensitive to make worthwhile comparisons with C$_3$H. 

\end{itemize}

The situation is summarized in Figure 5. At left, the molecular column
densities are plotted against the far IR dust emission-derived optical 
reddening equivalents given in Table 1 \citep{SchFin+98}
and only the total column density toward B0355 can
be used; at right the individual kinematic components have been 
measured for several but not all molecules toward B0355.  Also
shown in this Figure are values of N(C$_3$) and N(C$_2$), using the
C$_3$ column densities of \cite{AdaBla+03} and \cite{OkaTho+03}
\footnote{The C$_3$ column densities in common between these references 
only agree to within a factor two or so}, the C$_2$ column densities
cited in either paper and the CH column densities given by 
\cite{OkaTho+03}. 
Inclusion of the results for C$_2$ and C$_3$ was motivated by 
the central role attributed to C$_3$ in small-hydrocarbon formation 
in dark clouds by \cite{LoiAgu+17}, see their Figure 3. Ironically,
N(C$_3$) is not observable in dark clouds so \cite{LoiAgu+17} did not 
tabulate calculated values of N(C$_3$) from their models.  Triatomic
carbon is observed at THz frequencies in the envelopes and cores
of star-forming regions like DR21(OH) with a fractional abundance 
X(C$_3$) $\approx 0.6 - 3.0 \times 10^{-9}$ \citep{MooHas+12}, 
comparable to what is shown here in Figure 5 \footnote{Abundances
of \cch\ and \cyclic\ are also comparable.}.  CO aside, C$_2$ 
is the most abundant carbon-bearing molecule in diffuse molecular 
gas, 2-3 times more abundant than either CH or \cch.  The factor 
40 drop in abundance between C$_2$ and C$_3$ is twice as 
large as that between \cch\ and \cyclic.

\cite{LoiAgu+17} consider in detail the formation of the isomers
of the molecular ions that recombine to form C$_3$H and C$_3$\HH\ 
along with those of their recombination products, and they took into
account the subsequent linear $\rightarrow$ cyclic isomerization 
arising from reaction with free atomic hydrogen.  They conclude 
that the comparatively small $c$-C$_3$H/$l$- C$_3$H ratio 
($\approx 5$) in dark clouds is created during initial formation, 
either via the reaction of C + C$_2$\HH\ or by dissociative recombination, 
while the much larger values 30-100 seen in C$_3$\HH\ arise after 
formation of \linear\ through linear $\rightarrow$ cyclic isomerization 
in reaction with atomic hydrogen.  

The $c$-C$_3$\HH/$l$-C$_3$\HH\ ratio in dark clouds decreases 
with increasing density, which is understood in terms of the 
smaller atomic hydrogen fraction in denser gas.  While this
may occur, the much larger atomic hydrogen fraction 
in diffuse and translucent gas does not lead to yet-larger
$c$-C$_3$\HH/$l$-C$_3$\HH\ ratios in our observations, which
show quite comparable values to those seen in dark clouds.  
The inverted ratios  $c$-C$_3$H/$l$-C$_3$H $\approx 0.5$ in 
our work have no precedent in dark clouds.

\begin{figure}
\includegraphics[height=5cm]{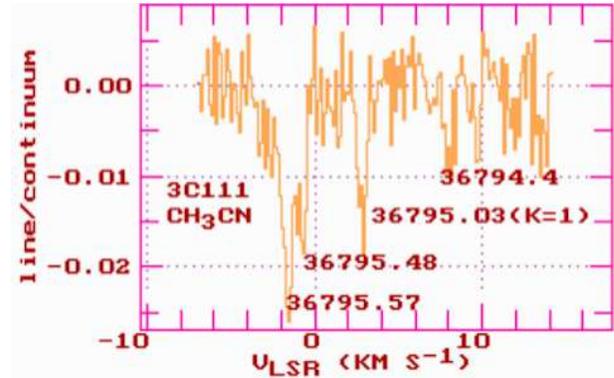}
  \caption{ Closeup of \methCN\ K=0 and K=1 lines (Table 1) toward B0415 at 0.127 \kms\ spectral
 resolution.}
\end{figure}

\section{The abundance of polyynes and CN-bearing species}

\begin{itemize}

\item CN itself is the most abundant CN-bearing molecule. 
 N(CN)/N(\HH) $= 3\times 10^{-8}$ in diffuse molecular gas and toward TMC-1,
 or  N(CN)/N(\HH) $= 1\times 10^{-8}$ toward B1b.

\item N(CN)/N(HCN) $\approx$ 7 in diffuse molecular gas vs. 1-1.5 in dark cloud gas

\item N(HCN)/N(HNC) $ \approx 3-6$ in diffuse molecular gas vs. 1 in dark cloud
gas, a sign of warmer chemistry in diffuse gas.

\item N(HC$_3$N)/N(HCN) $\le 0.4$ in diffuse molecular gas comparable to B1b but
much less than TMC-1 where N(HC$_3$N)/N(HCN) $\approx 60$.

\item N(\methCN)/N(HCN) $\approx 0.015$ in diffuse molecular gas comparable to 
TMC-1 (0.02) but much greater than B1b (0.002).

\item N(\methCN)/N(HC$_3$N) = 4 toward B0415, much larger than in dark clouds
(1/20 - 1/30), so \methCN\ is enhanced but not by as much as in the Horsehead
PDR.

\item The large values N(C\HH CN)/N(\methCN) $\approx 10$ in dark clouds are 
not seen in diffuse molecular gas.

\item There is no fiducial value for N(CH$_3$NC)/N(\methCN) in dark cloud gas
but the best upper limits N(CH$_3$NC)/N(\methCN) $< 0.15 - 0.3$ in diffuse molecular 
gas are comparable to the abundance ratio N(CH$_3$NC)/N(\methCN) $= 0.15$ seen toward
the Horsehead PDR.

\end{itemize}

The overall situation is summarized in Figure 6 where the optical 
absorption measurements of N(CN) and N(C$_2$) cited by 
\cite{OkaTho+03} are also included.  CN itself is the most
abundant CN-bearing molecule, with column densities about 
1/3 - 1/2 those of C$_2$, or comparable to those of CH, at larger 
\EBV\ or N(CH).  The main result is of course the surprising 
ubiquity of \methCN\ in diffuse molecular gas, with X(\methCN) 
= N(\methCN)/N(\HH) $\approx 0.85 \times 10^{-10}$. That said, there 
is another surprise in Figure 6:  optical CN absorption line data 
at intermediate \EBV\ or N(CH) where N(CN) measured in optical 
absorption is much smaller than N(CN) measured in the radio at the same 
\EBV.  In mm-wave absorption, CN, HCN and HNC appear in nearly 
fixed proportions \citep{LisLuc01,AndKoh+16}, 
with N(CN)/N(HCN) = $7\pm 1$. Smaller CN abundances measured 
in absorption toward early-type stars would suggest photodissociation 
of CN, especially, as the likely cause.
Lamentably, the abilities of optical/UV absorption spectroscopy have 
not yet allowed detection of species such as HCN in the absorption
spectra of stars occulted by diffuse clouds.  The optical/UV spectra 
of HCN and HNC have  recently been calculated by \cite{AguRon+17} 
as part of  a computation of the photodissociation rates of both 
species, showing that the photodissociation rate of HNC is 2.2 times 
greater.  This could account in part for the higher N(HCN)/N(HNC)
ratios in diffuse clouds, compared to TMC-1 (see Figure 6 and Table
5).

\section{Oxygen-bearing species}

Limits on the column densities of isocyanic acid (HNCO), formic acid 
(HCOOH; found on Earth in ants, bees and nettle plants according to its 
discoverers \cite{ZucBal+71}) in the interstellar medium (ISM) 
and protonated formaldehyde (\HH COH\p) are given in Table 6.  HNCO and HCOOH 
were observed in their lowest transitions, leading to rather large 
uncertainties in their column densities as noted in Table 1.  HNCO can only 
be said to be less abundant in diffuse molecular gas than in TMC-1 if the 
excitation is weak in the diffuse molecular gas.  There is no fiducial value 
of the column density of protonated formaldehyde (\HH COH\p) for TMC-1.

\section{C\HH CN\m\ as a possible DIB carrier}

\begin{figure*}
\includegraphics[height=8.4cm]{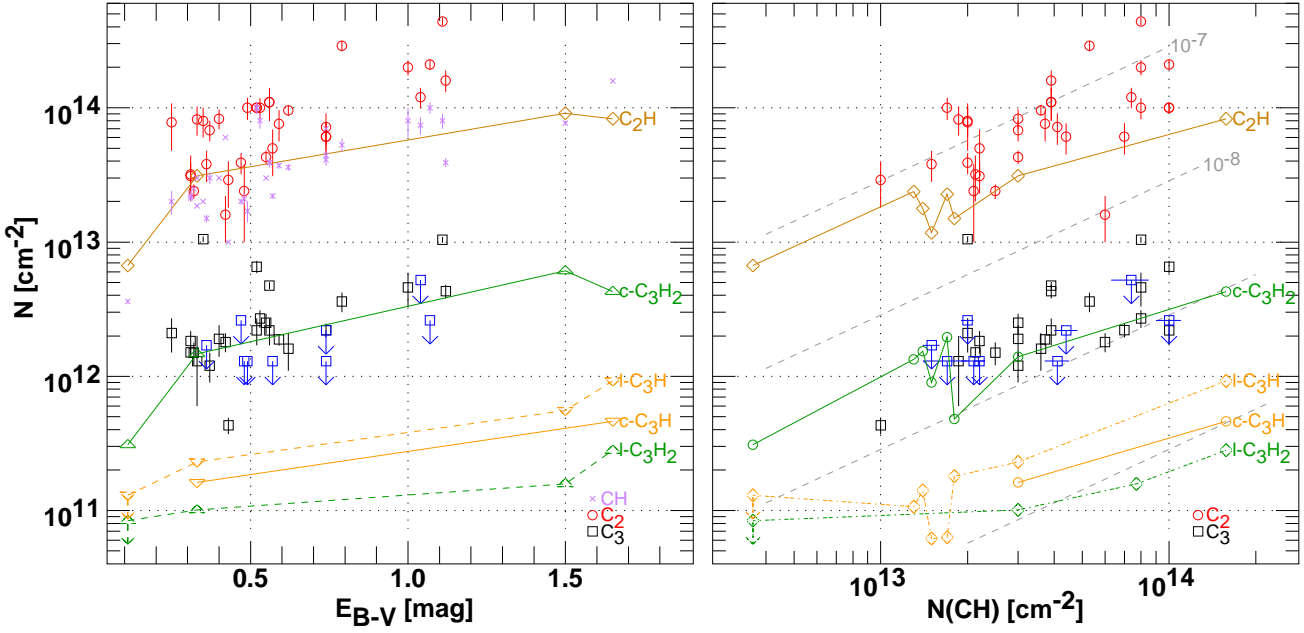}
  \caption{Column densities of C$_3$, C$_2$ 
\citep{AdaBla+03,OkaTho+03}, CH \citep{OkaTho+03} and small 
hydrocarbons (Table 4). 
Shown at left are column densities plotted against the IR 
dust emission-derived optical reddening equivalents (Table 1) for
the radio data, or using the stellar reddening for the optical C$_2$,
C$_3$ and CH data. At right, column densities are plotted against 
N(CH) using the individual component column densities for the
radio data where possible and using N(CH) cited by \cite{OkaTho+03}
for the sightlines observed in optical absorption.  Dashed gray lines
at right show fractional abundance with respect to \HH\ assuming
N(CH)/N(\HH) $= 3.5\times 10^{-8}$.}
\end{figure*}

\cite{CorSar07} proposed the para-ladder rotational transitions of
the anion C\HH CN\m\ as the carrier of a diffuse interstellar band 
(DIB) at $\lambda$803.7nm. It is this ladder whose lowest rotational 
transition was observed here in the neutral version of the molecule, 
C\HH CN.  The neutral and its anion have the same symmetry properties,
similar rotational structure and  roughly comparable permanent dipole 
moments (3.6 vs. 1.2 Debye, respectively) \citep{MajDas+14}
 so that arguments used in the discussion of the required
abundance of the anion 
should also be used when comparing its column density with that of the 
neutral observed here. As shown in Table 3, the ratios N(C\HH CN)/N(CN) 
and N(C\HH CN)/N(HCN) are at least about one order of magnitude smaller 
in diffuse molecular gas than in dark cloud gas toward TMC-1.

Using Herbig's unpublished data \cite{CorSar07} determined equivalent 
widths toward eight stars having reddening \EBV\ $= 1 - 1.4$ mag, quite 
comparable to those toward B0355 and B0415 in this work.  The results
are $<W_\lambda$/\EBV$> = 0.00220\pm0.00064$ nm/mag, or
$<$ N($p$-C\HH CN\m)/\EBV $> = 4.025\times 10^{10}/{\rm f}\pcc$/mag
where f is the oscillator strength of the $\lambda$803.7nm transition.
\cite{CorSar07} hypothesized f=0.5, leading to an implied column
density N($p$-C\HH CN\m) $= 1.2-1.3\times 10^{11}\pcc$ toward B0355
and B0415.  The upper limits we deduce for N($p$-C\HH CN) toward these 
sources are somewhat above this, N($p$-C\HH CN) $< 2.5\times 10^{11}\pcc$
before applying a correction for rotational excitation above that 
provided by radiative equilibrium with the cosmic microwave background, 
which is in the range 1-3.

\cite{CorSar07} showed that two strong spectral features corresponding to 
absorption out of the ortho-ladder K=1 levels were absent in the optical 
spectrum, implying that all of the C\HH CN\m\ resided in the para 
rotational ladder\footnote{In fact this could easily be taken to 
disqualify C\HH CN\m\ as the carrier.}.
To explain this, \cite{CorSar07} argued that the ortho/para ratio 
was small because weak collisional excitation in the diffuse molecular
ISM would leave all molecules in the lowest possible states, in
radiative equilibrium with the cosmic microwave background in all
facets of the excitation.  Our calculations show that this
is a poor assumption for the para-ladder given the large electron 
fraction in diffuse molecular gas and the large permanent dipole
moments of the species in question, but the optical profiles that 
were integrated to give the equivalent widths naturally include the 
poorly-resolved rotational sub-structure even if \cite{CorSar07} did 
not consider it to be present.  The point is that we are obliged to compare the 
required column density of $p$-C\HH CN\m\ with upper limits for 
$p$-N(C\HH CN) that are fully corrected for rotational excitation 
within the para-rotation ladder even if they weaken our conclusions.

Our limits on N($p$-C\HH CN) are above the required column density of the 
anion by a factor of a few, 2-6.  Under normal circumstances, the large 
neutral/anion column density ratios $> 200$ found for other species 
\citep{SatGia+15} would exclude C\HH CN\m\ as a possible carrier 
of the DIB at $\lambda$803.7nm.  However, \cite{CorSar07} argued, 
on the basis of unpublished work by E. Herbst and T. Millar, that the 
neutral/anion ratio would be exceptionally small, N(C\HH CN)/N(C\HH CN\m) 
$\approx 1$. 

Indeed, small ratios N(C\HH CN)/N(C\HH CN\m) = 0.25 - 0.6
were subsequently calculated by \cite{MajDas+14} who tracked the time
evolution of a comprehensive chemical network over a wide 
range of \AV\ and n(H).  However, the models of \cite{MajDas+14} 
also predict N(C\HH CN\m) $ = 3.5\times 10^7\pcc$  and 
N(\methCN) $= 1.4 \times 10^5\pcc$ at \AV\ = 1 mag and 
n(H) $=350 \pccc$.  These are some 4 orders of magnitude below
the required column density of N(C\HH CN\m) but also more than
five orders of magnitude below our newly-observed column density of 
\methCN\ toward B2200+420 (\bll) at \AV\ = 1 mag in Table 5.
Clearly, the chemistry of C\HH CN\m, and other important anions 
and molecules possibly linked to DIBs in diffuse molecular gas, must 
be revisited.

To summarize, our observational upper limit suffices to
show that the ratios N(C\HH CN)/N(CN) and N(C\HH CN)/N(HCN) are at 
least about one order of magnitude smaller in diffuse molecular gas 
than  toward TMC-1.  But if it is accepted that the neutral/anion 
ratio is so much smaller for C\HH CN than for other species, 
C\HH CN\m\ might remain a viable carrier of the DIB at 
$\lambda$803.7nm.

\section{COMS in diffuse clouds?}

Claims for the presence of various oxygen and 
nitrogen bearing complex organic molecules (COMS) in diffuse clouds
have recently been made on the basis of ALMA observations toward Sgr 
B2 \citep{ThiBel+17}.  Some of the column densities derived in that 
work are shown in Table 7, where we copied results  for the three 
galactic center clouds appearing near 0-velocity (their Table 1)
and for the cloud at +27 \kms\ assumed to lie in the Scutum arm (their
Table 2).  For comparison we show results for TMC-1 (\AV\ = 10-20 mag)
and B2200 (\AV\ = 1 mag),
largely as shown in our Tables 5 and 6.  For TMC-1 and B2200 we take 
N(H$^{13}$CO\p) = N(\hcop)/62, the result obtained for local
gas \citep{LucLis98}.  The results for N(CH$_3$OH) are taken from 
from \cite{LisPet+08} for B2200 and from \cite{OhiIrv+92} and 
\cite{GraMaj+16} for TMC-1.

\hcop and \cyclic\ are often used as \HH\ tracers, for instance
with X(\hcop) = N(\hcop)/N(\HH) $= 3\times 10^{-9}$ here, or
X(\cyclic) $= 2.5\times 10^{-9}$ in the work of \cite{RiqBro+17}.
As shown in Table 7, the column densities of \hcop\ in the features 
described as diffuse clouds toward Sgr B2 range from 15 to 200 times 
larger than toward B2200 and are comparable to or even larger than what 
is observed in TMC-1\footnote{The very largest disparities might
be explained in small part by a smaller N(\hcop)/N(H$^{13}$CO\p) ratio
if the material near 0-velocity toward Sgr B2 is actually in the central 
molecular zone.}.  The \cyclic\ column densities seen toward Sgr B2 range
up to 12 times that seen toward B2200.  Clouds with such comparatively
high column densities of the tracers of \HH\ cannot also have 
\AV $\la$ 1 mag, the usual meaning of the term ``diffuse'' \citep{SnoMcC06}.

The column densities toward Sgr B2 are 3-20 times larger than toward 
TMC-1 for CH$_3$OH, 3-5 times larger than TMC-1 for CH$_3$CN, and as 
much as 5 times larger than TMC-1 for HC$_3$N.  They are all several 
hundred times larger than seen toward B2200.  COMS may have been
observed toward Sgr B2, but the nature of the host gas remains to
be determined.


\begin{figure*}
\includegraphics[height=9.1cm]{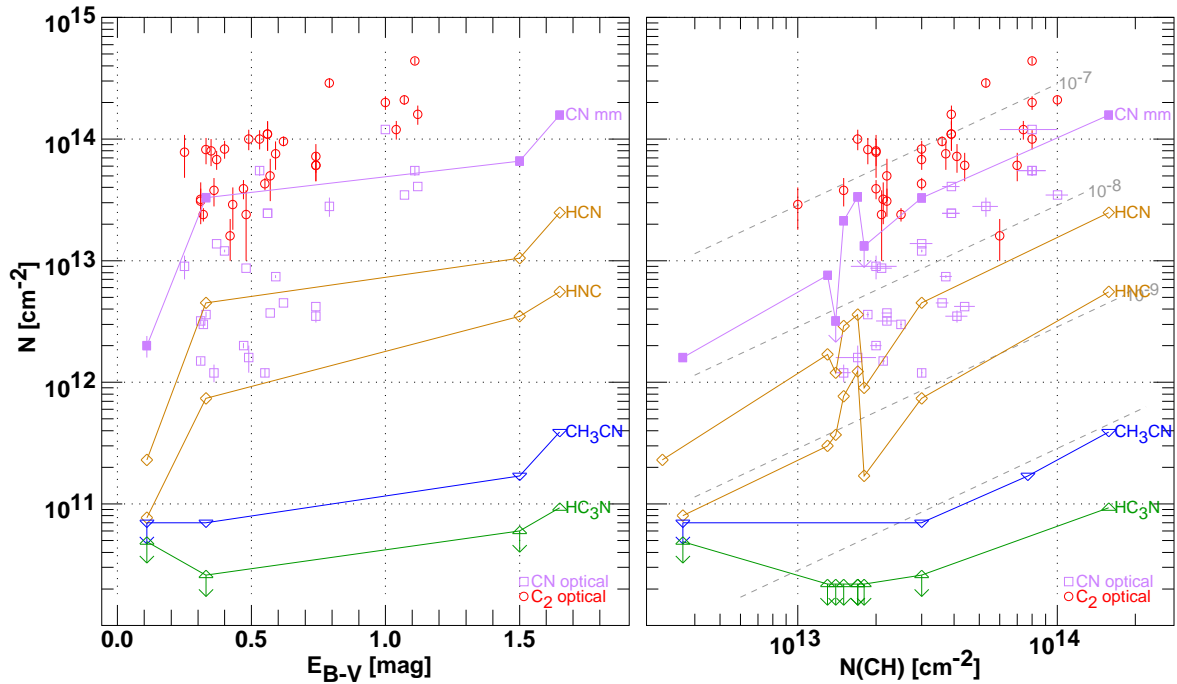}
\caption{Column densities of C$_2$ \citep{AdaBla+03,OkaTho+03}, 
CH \citep{OkaTho+03} and CN-bearing species (Table 5). 
Shown at left are column densities plotted against the IR 
dust emission-derived optical reddening equivalents (Table 1) for
the radio data, or using the stellar reddening for the optical C$_2$
and CN data. At right, column densities are plotted against 
N(CH) using the individual component column densities for the
radio data where possible and using N(CH) cited by \cite{OkaTho+03}
for the sightlines observed in optical absorption.  Dashed gray lines
at right show fractional abundance with respect to \HH\ assuming
N(CH)/N(\HH) $= 3.5\times 10^{-8}$.}
\end{figure*}

\section{Summary and discussion}

This work completes several major aspects of a long work program to 
catalog and systematize the molecular inventory of diffuse molecular 
gas observed in absorption at radio wavelengths near the Sun and 
in the wider Galaxy outside the central molecular zone, extending it 
beyond the very limited complement of mostly-diatomic molecules seen at 
UV through NIR wavelengths.  The case for comparability of the 
diffuse molecular gas observed in the radio and UV through NIR domains 
was made in our recent discussions of the suitability of small polar
species as carriers of DIBs \citep{LisSon+12,LisLuc+14} and will not be
repeated here, keeping the focus on the observable chemistry of the
detected hydrocarbons and CN-bearing species. The oxygen-bearing
family of molecules observed at radio wavelengths 
(OH, CO, HCO, \hcop, HOC\p, \HH CO and CH$_3$OH) will be discussed 
in a forthcoming work that includes recent ALMA observations of 
HOC\p\ and comparisons with existing HERSCHEL observations of \HH O.   

The systematics of the small hydrocarbons and CN-bearing species are 
comprehensively outlined in Sections 3 and 4, respectively. The most 
abundant species in each family, CH or \cch\ and CN, have relative 
abundances with respect to \HH\ that are about equal to each other 
and the same in diffuse molecular gas and TMC-1, X(\cch) = N(\cch)/N(\HH) 
$= 4\times 10^{-8}$ and X(CN) $= 3\times 10^{-8}$ at higher \EBV\ or 
N(\HH).  However, the most abundant carbon-bearing molecule overall among 
those considered here (ie, neglecting CO), is C$_2$ with 
X(C$_2$) $= 8\times 10^{-8}$, 2-3 times more abundant than CH, 
\cch\ or CN.  The factor 40 drop in abundance between C$_2$ and C$_3$ is 
twice as large as that between \cch\ and \cyclic.

In this work we showed that $l$-C$_3$H and \methCN\ are ubiquitous
in local diffuse molecular gas and the ratio of \methCN\ to 
HCN is the same as in TMC-1, N(\methCN)/N(HCN) $\approx 0.02$.  
The relative
abundance of $c$-C$_3$H is about the same in diffuse molecular gas
as in TMC-1 or dark clouds generally \cite{LisPet+14} but the linear
variant is enhanced in diffuse molecular gas:
N($c$-C$_3$H)/N($l$-C$_3$H) $\approx 0.5$ in diffuse molecular gas,
vs 4-10 in the other environments considered in Tables 3-4.  The linear
variant is  much more abundant relative to cyclic in C$_3$H than in
C$_3$\HH\ in all environments.  

The $c$-C$_3$\HH/$l$-C$_3$\HH\ ratio in dark clouds decreases 
with increasing density, which is understood in terms of the 
smaller atomic hydrogen fraction in denser gas.  The much larger 
atomic hydrogen fraction in diffuse and translucent gas does not 
lead to yet-larger $c$-C$_3$\HH/$l$-C$_3$\HH\ ratios in our observations, 
which show quite comparable values to those seen in dark clouds.  
The inverted ratios $c$-C$_3$H/$l$-C$_3$H $\approx 0.5$ in our work 
have no precedent in dark clouds.

In Section 6 we discussed the suitability of C\HH CN\m\ as a DIB-carrier
\citep{CorSar07} based on the limits we were able to set on N(C\HH CN).  
For C\HH CN to be a viable candidate DIB-carrier, the neutral/anion
ratio would have to be small, no more than 2-6. Neutral/anion
ratios for observed species are typically 200:1 or larger \citep{SatGia+15}.

In Section 7 we compared our results with those of \cite{ThiBel+17} for
three low-velocity clouds and another in the Scutum Arm observed
in absorption toward Sgr B2:  these observations are the basis of claims
for the existence of complex organic molecules (COMS) in diffuse clouds.   
We noted that column densities of \HH-tracers such as \hcop\ were one - two 
orders of magnitude higher in that work than those we associate with clouds at 
\AV\ $\la$ 1 mag locally, and in some cases even larger than those seen in TMC-1.  
Claims for the presence of COMS in diffuse clouds, material at \AV\ $\la$ 1 mag, 
must be carefully assessed.

Describing the molecular inventory of diffuse molecular gas is still a work 
in progress: outstanding undetected hydrocarbons with three carbon atoms 
include \methCCH\ and the recently-introduced t-C$_3$\HH\ \citep{LoiAgu+17} 
whose microwave spectrum is unknown. t-C$_3$\HH\ could be a common host of 
unidentified lines given the ubiquity of the other isomers of C$_3$\HH\ 
in a wide range of astrophysical environments. 

Understanding the observed abundance 
of even some quite small species (CH\p, \hcop) in diffuse molecular gas 
requires the addition of new physics into the chemical modelling, as 
embodied in the work of \cite{GodFal+14} and \cite{ValGod+17}.  It has 
further been suggested that the small hydrocarbons observed here should
originate in a top-down chemistry after  the breakup of much larger species 
\citep{GuzPet+15}.  Observations of polycyclic aromatic hydrocarbons 
with the James Webb Space Telescope may soon  test this idea.

\acknowledgments

  The National Radio Astronomy Observatory is operated by Associated
  Universities, Inc. under a contract with the National Science Foundation.
  HL, MG and JP were partially funded by the grant ANR-09-BLAN-0231-01 from the 
  French {\it Agence Nationale de la Recherche} as part of the SCHISM 
  project (http://schism.ens.fr/) during the early phases of this work. The work 
  of MG and JP was supported by the CNRS program ``Physique et Chimie du 
  Milieu Interstellaire''(PCMI). The work of MG and JP was supported by the 
  Programme National ``Physique et Chimie du Milieu Interstellaire'' (PCMI) of 
  CNRS/INSU with INC/INP co-funded by CEA and CNES.

  We thank the Alexandre Faure for providing excitation rates for \methCN\
  and we thank the anonymous referee for a variety of remarks that led to 
  improvements in the manuscript.

\facility{VLA}

\software{DRAWSPEC (Liszt 1997), CASA (McMullin et al. 2007)}




\begin{table}
\caption{Continuum target and sightline  properties}
{
\small
\begin{tabular}{lccccc}
\hline
Target & aka &  l    &  b    & \EBV$^a$ & flux$^b$  \\
       &     & \degr & \degr & mag      &  \% \\
\hline
B0355+508 & NRAO150 & 150.38 &-1.60 & 1.50 & 28,29  \\
B0415+379 & 3C111  & 161.67& -8.82 & 1.65 & 8,10 \\
B2200+420 & \bll & 92.59& -10.44 & 0.33 & 28,31 \\
B2251+158 & 3C454.3 & 86.11& -38.18 &0.11 & 16,21  \\
\tableline
\end{tabular}}
\\
$^a$from \cite{SchFin+98} \\
$^b$ entries are 21 GHz and 36 GHz fluxes as percentages \\
of 3C84 (S$_\nu \approx 16,10$ Jy)\\
\end{table}

\begin{table*}
\caption{Species and transitions observed and column density-optical 
depth conversion factors at T$_{\rm ex}$=T$_{\rm CMB}$}
{
\small
\begin{tabular}{lcccccc}
\hline
Species & ortho/para/other & transition & frequency& log(A$_{kj}~\ps)^a$ & N(X)/$\int\tau dv^b$ & Correction$^e$\\
        &   &                        & MHz  &     & $\pcc$ (\kms)$^{-1}$ & \\
\hline
\linearC3H & $l=f$ & J=3/2-1/2,$\Omega$=1/2,F=2-1 & 32627.30 & -5.89 & $2.82\times10^{13} $ & 1-1.8\\
\linearC3H & $l=e$ & J=3/2-1/2,$\Omega$=1/2,F=2-1 & 32660.65 & -5.89 & $2.82\times10^{13} $ & 1-1.8\\
HC$_3$N$^c$  &  & J=4-3 & 36292.33 &  -5.49 & $1.09\times10^{13}$ & 1-1.6 \\
CH$_3$CN  & E  & 2(0)-1(0) F=3-2 & 36795.57 &  -5.45 & $1.27\times10^{13}$  &  1-1.7\\
CH$_3$CN  & E  & 2(0)-1(0) F=2-1 & 36795.48 &  -5.57 & $2.38\times10^{13}$  & 1-1.7  \\
CH$_3$CN  & E  & 2(0)-1(0) F=1-0 & 36794.42 &  -5.70 & $5.47\times10^{13}$  & 1-1.7  \\
CH$_3$CN  & A  & 2(1)-1(1) F=1-0 & 36795.03 &  -5.57 & $1.19\times10^{13}$  & 1-1.7 \\
CH$_3$NC  & E   & 1(0)-0(0) & 20105.75 & -6.32 & $1.07\times10^{13}$  & 1-4.5 \\
C\HH CN$^d$ & p  & $1_{01}-0_{00}$ & 20119.61 & -6.41 & $7.73\times10^{13}$ & 1-3.5 \\
HNCO        &  & 1(0,1)-0(0,0) F=2-1  & 21981.46 & -6.98 &  $6.39\times10^{13}$ & 1-6 \\
HCOOH       & t& 1(0,1)-0(0,0)   & 22471.18 & -7.07 &  $6.80\times10^{13}$  & 1-7 \\
\HH COH\p   & & 2(0,2)-1(1,1)   & 36299.95 & -6.51 &  $5.61\times10^{13}$  & 1-1.6 \\
\tableline
\end{tabular}}
\\
$^a$ www.splatalogue.net \\
$^b$ for the observed ortho or para version only, assuming rotational \\
excitation in equilibrium with the cosmic microwave background \\
$^c$ 96\% of the integrated intensity is in an unresolved blend \\
$^d$ J=3/2-1/2,F$_1$=5/2-3/2,F=7/2-5/2. Spectroscopy from \cite{IrvFri+88} and \cite{OhiKai98} \\
$^e$ See Figures A1-A2 \\
\end{table*}

\begin{table*}
\caption{Integrated optical depths (EW) for newly-observed species  $^a$}
{
\small
\begin{tabular}{lcccccccccc}
\hline
Target &  vel  & EW & EW  & EW  & EW  & EW & EW & EW & EW & EW \\
       &\kms  & m \ps & m \ps   & m \ps    & m \ps & m \ps & m  \ps  & m  \ps  & m  \ps  & m  \ps    \\
\hline
 &        & \linearC3H & HC$_3$N & CH$_3$CN$^b$ & CH$_3$CN$^c$ &  CH$_3$NC & C\HH CN &HNCO&HCOOH&\HH COH\p\\
\hline
B0355+508 & -17 &  $2.20(0.56)$ & $< 2.0$ & & & $<1.71$  & $<$ 1.90 &$<$2.10&$<$2.21&$<$3.63\\
   & -14  & $6.40(0.64)$ &  & & & & &&&\\
  & -10 &  $2.22(0.48)$&  & & & & &&&\\
  & -8 &  $3.56(0.56)$ & & & & &&& &\\
  & -4 &  $5.00(0.70)$ &  & & & &&&&\\
   & all &  $19.3(0.13)$ & $< 5.5$ & 23.0(3.0)$^d$ & & $<4.47$ & $<$3.24 &$<$4.86&$<$4.80&$<$7.98 \\
\hline
B0415+379 & &  $32.8(1.75)$ &  8.5(1.9) & 42.4(3.2) & 9.7(1.8) & $<4.16$ & $<$3.15&$<$4.62&$<$4.62&$<$9.30 \\
B2200+420 & & $8.16(0.4)$ & $<2.55$ &7.4(1.2)  &2.3(0.6) & $<1.92$ &$<$1.68&$<$1.65&$<$1.95&$<$2.94 \\
B2251+158 & & $< 4.5$ & $< 4.5$ & $<$7.1 & & $<4.92$ & $<$3.90&&&$<$7.31 \\
\tableline
\end{tabular}}
\\
$^a$ all upper limits are $3\sigma$ \\

$^b$ The sum of the three observed K=0 lines \\
$^c$ K=1 \\
$^d$ K=0 and K=1 are not distinguishable, this is their sum \\
\end{table*}
 
\begin{table*}
\caption{Column densities for hydrocarbons}
{
\small
\begin{tabular}{lccccccccc}
\hline
Target & v & N(\HH)$^1$ & N(CH)$^2$ &N(\cch)$^a$ & N(\cyclic)$^b$ & N(\linear)$^c$  & N(C$_4$H)$^m$ & N(\cyclicC3H)$^d$ & N(\linearC3H)  \\
       & \kms\  & $10^{20}\pcc$  &  $10^{13}\pcc$ &  $10^{13}\pcc$ & $10^{12}$ $\pcc$ & $10^{11}\pcc$  & $10^{13}\pcc$ & $10^{11}\pcc$ & $10^{11}\pcc$ \\ 
\hline
B0355  & -17  &4.3  &1.5 & 1.17  & 0.90  & &&& 0.62(0.16)  \\
       & -14  & 5.0 &1.8 & 1.50  & 0.48 & &&& 1.81(0.18)   \\
       & -10  & 4.8 &1.7 & 2.27& 1.96 & &&& 0.63(0..14)  \\
       & -8   & 3.8  &1.3 & 2.38 & 1.34 & &&& 1.07(0.16)  \\
       & -4   & 4.0 &1.4 & 1.78 & 1.54 & &&& 1.41 (0.20) \\
       & -all & 22 & 7.7& 9.10 & 6.11  &  1.58  & $<$1 & & 5.56(0.34) \\
\hline
B0415 &    & 45 &15.8& 8.29 & 4.28& 2.81 &  $<$ 2.3 & 4.63(0.18)  & 9.24(0.49) \\
B2200 &  & 8.7  &3.0& 3.11 & 1.47& 1.01   & $<$ 0.4  & 1.62(0.05) & 2.30(0.13) \\
B2251$^e$ &  & 1.0 &0.36 & 0.67 & 0.31  & $<$ 0.84  &$<$ 0.3 & & $<$ 1.3 \\
\hline
TMC-1/10$^f$ &  &10& 2 & 5-10  & 10  &   & 2    &  6  & 5  \\
TMC-1/10$^g$ &  &'' & &       & 2   & 0.6  & 0.3-9 & 18  & 6\\
TMC-1/10$^h$ &  &''& & 6     & 12   & 2 &   & 10   & 1 \\
TMC-1/10$^i$ &  &''& & 2     & 6   & 2 &   & 10  & 1 \\
consensus    & &10-20$^l$ &2 &5 & 6 & 1 & 2 & 9 & 2  \\
\hline
B1b/10$^j$ & &$\ga60$  & &   & 2   & 0.6 & & 6   & 1  \\
\hline
HH PDR/10$^k$  &   &19  &      & 1-2  & 0.5-0.8 & 0.5-1.5&  & 2-7    & 0.6 - 1.8  \\
HH core/10$^k$  &  &32  &      & $<1$ & 0.3-0.4 & 0.1-0.3 & & 0.8-2.3& 0.1-0.4  \\
\hline
Orion Bar/10$^n$&  & 30   &      & 4 &  1.3 & 0.4 & 0.4 & 2 & 0.6 \\
\tableline
\end{tabular}}
\\
$^1$ N(\HH) = N(\hcop)/$3\times 10^{-9}$ for sources observed in this work \\
$^2$ N(CH) = N(\HH) $\times 3.5 \times 10^{-8}$ for sources observed in this work \\
$^a$N(\cch) from \cite{LucLis00} \\
$^b$N(\cyclic)= $(4/3)\times$N($o$-\cyclic) from \cite{LisSon+12}\\
$^c$N(\linear)= $4\times$N($p$-\linear) from \cite{LisSon+12}\\
$^d$ N($c$-C$_3$H) from \cite{LisPet+14}\\
$^e$ upper limits are $3\sigma$ \\
$^f$ \cite{OhiIrv+92} whose tables must be interpreted with N(\HH) $= 10^{22}\pcc$ \\
$^g$ \cite{GraMaj+16}\\
$^h$ \cite{LoiAgu+17} except \cch\ from \cite{SakSar+10}  \\
$^i$ \cite{FosCer+01} \\
$^j$ \cite{LoiAgu+17} and \cite{DanGer+13} \\
$^k$ Horsehead (HH nebula values from \cite{GuzPet+15} \\ 
$^l$ N(\HH) $\ge 2\times 10^{22}\pcc$ beam-averaged on ~ 1\arcmin\ scales 
 is given by \cite{FehTot+16} \\
$^m$ Results for C$_4$H from \cite{LisSon+12} \\
$^n$ \cite{CuaGoi+15}, Table 6 \\
\end{table*}

\begin{table*}
\caption{Column densities for CN-family molecules$^a$}
{
\small
\begin{tabular}{lccccccccc}
\hline
Target & v & N(\HH) & N(CN)$^b$ & N(HCN)$^b$  & N(HNC)$^b$ & N(HC$_3$N) &N(\methCN)$^{c}$ & N(\methNC) & N(C\HH CN)  \\
       & \kms\ & $10^{20}\pcc$& $10^{13}\pcc$ & $10^{13}\pcc$ & $10^{13}\pcc$ & $10^{11}\pcc$ & $10^{11}\pcc$& $10^{11}\pcc$ & $10^{11}\pcc$ \\
\hline
B0355  & -17  &4.3 &2.13&  0.29 &0.077 &$<$0.22&  & $<$ 0.18 & $<$  1.5  \\
       & -14  &5.0&$<$0.32& 0.09  &0.017& &  & & \\
       & -10  &4.8&3.35 &   0.36 &0.123 & & & & \\
       & -8   &3.8&0.76 &0.17 &0.030& & &  &\\
       & -4   &4.0&$<$0.32&0.12 &0.037 & & & & \\
       & -all &22&6.6&1.05 & 0.28&  $<$0.60      &1.7(0.2) & $<$0.37&  $<$  2.5\\
\hline
B0415  & &45 &15.78& 2.480& 0.554& 0.93(0.21)&3.9(0.3) & $<$0.44&  $<$ 2.4 \\
B2200 &  & 8.7& 3.29 &0.450 & 0.074 & $<$0.26&0.7(0.1) & $<$0.20 &  $<$ 1.3  \\ 
B2251 & & 1.0 & 0.20 &0.023 &0.008 &$<$0.49 & $<$0.7   & $<$0.52 &  $<$ 3.0  \\
\hline
TMC-1/10$^d$ & & 10 &3 & 2 & 2 & 60  & 10 & & 50  \\
TMC-1/10$^e$ & &  &  &   &   & 234 & 4  & & 38  \\
consensus   & &  $10-20^i$ &3 &2  &2  &120 & 6   & &44 \\
\hline
B1b/10$^g$ &&$\ga60$&6&5&2&2&0.1&& \\
\hline
HH-PDR$^h$  &&&&&& 2.5&100&15& \\
HH-core$^h$ &&&&&& 5  &5  &$<$5& \\
\hline
Orion Bar$^j$/10 & &30  & 2.5  & 0.34   & 0.4   & 3 & 7  & &   \\
\tableline
\end{tabular}}
\\
$^a$ all upper limits are $3\sigma$ \\
$^b$N(CN), N(HCN) and N(HNC) from \cite{LisLuc01} \\
$^c$Sum of N(\methCN) K=0 and K=1 \\
$^d$ \cite{OhiIrv+92} whose tables must be interpreted with N(\HH) $= 10^{22}\pcc$ \\
$^e$ \cite{GraMaj+16} \\
$^g$ \cite{LoiAgu+17} and \cite{DanGer+13} \\
$^h$Horsehead nebula values from \cite{PetGra+12}, \cite{GraPet+13} and \cite{GuzPet+15} \\
$^i$ N(\HH) $\ge 2\times 10^{22}\pcc$ beam-averaged on ~ 1\arcmin\ scales 
is given by \cite{FehTot+16} \\
$^j$ \cite{CuaGoi+17}

\end{table*}

\begin{table}
\caption{Column densities for oxygen-bearing molecules$^a$}
{
\small
\begin{tabular}{l c c c c c}
\hline
Target & v & N(\HH) & N(HNCO)&HCOOH&\HH COH\p \\
       & \kms  & $10^{20}\pcc$  & $10^{11}\pcc$  & $10^{11}\pcc$  & $10^{11}\pcc$ \\
\hline
B0355  & -17 & 4.3 & $<$1.3 &$<$1.5&$<$2.0 \\
       & -14 & 5.0  &  &&\\
       & -10& 4.8   & &&\\
       & -8 & 3.8  & &&\\
       & -4 & 3.8  & &&\\
       & -all & 22 & $<$3.1 &$<$3.2&$<$4.5\\
\hline
B0415   & & 45& $<$3.0 &$<$3.1&$<$5.2\\
B2200 &  & 8.7 & $<$1.1 &$<$1.3&$<$1.6 \\ 
B2251 &  & 1.0 &  && $<$4.1\\
\hline
TMC-1/10$^d$ & & 10  & 2    &$< 2$ & \\
TMC-1/10$^e$ & &     & 11   &    &   \\
consensus & & 10-20$^f$   & 4.7 & $<2$   &\\
\tableline
\end{tabular}}
\\
$^d$ TMC-1 values from  \cite{OhiIrv+92} \\
$^e$ TMC-1 values from  \cite{GraMaj+16} \\
$^f$ N(\HH) $\ge 2\times 10^{22}\pcc$ beam-averaged on 1\arcmin\ scales 
according to \cite{FehTot+16} \\
\end{table}

\begin{table*}
\caption{Comparison with Galactic Center and Scutum Arm diffuse clouds of \cite{ThiBel+17}}
{
\small
\begin{tabular}{lcccccc}
\hline
Species & TMC-1&B2200&GC 1& GC 2 & GC 3&Scutum \\
& $\pcc$& $\pcc$& $\pcc$& $\pcc$& $\pcc$& $\pcc$ \\
\hline
H$^{13}$CO\p & 1.3$\times10^{12~a}$ &0.042$\times 10^{12~a}$&1.5$\times 10^{12}$&8$\times 10^{12}$&4$\times 10^{12}$&0.6$\times 10^{12}$ \\
\cyclic &2.0$\times 10^{13}$&0.150$\times 10^{13}$&0.5$\times 10^{13}$&2$\times 10^{13}$&1$\times 10^{13}$&0.8$\times 10^{13}$ \\
CH$_3$OH &0.2$\times 10^{14}$&$<0.005\times 10^{14}$&4$\times 10^{14}$&4$\times 10^{14}$&2$\times 10^{14}$&0.6$\times 10^{14}$ \\
CH$_3$CN &0.4$\times 10^{13}$&0.007$\times 10^{13}$&1$\times 10^{13}$&2$\times 10^{13}$&$<6\times 10^{13}$&1.4$\times 10^{13}$ \\
HC$_3$N &13$\times 10^{13}$&$<0.004\times 10^{13}$&$<3\times 10^{13}$&60$\times 10^{13}$&$<3\times 10^{13}$&$<2.5\times 10^{13}$ \\
\tableline
\end{tabular}}
\\
$^a$ N(H$^{13}$CO\p) = N(\hcop)/62 \\
\end{table*}

\appendix

\section{Rotational excitation}

For the low-lying transitions of heavier species observed 
in this work, collisional excitation redistributes the rotational 
population out of the lowest states, increasing the numerical factors that 
should be used to convert observed optical depths to column density. 
Collisions with electrons greatly dominate the excitation in diffuse
molecular gas where the CO abundance is small and C\p\ is the dominant carrier 
of carbon leading to an electron fraction 
n(e)/n(H) $\ga 1.4\times10^{-4}$ \citep{SofLau+04}. 
Excitation rates for collisions with He and \HH\ play a smaller role and
have not been calculated for most of the species discussed here but we 
included \HH\ excitation of HC$_3$N \citep{FauLiq+16} and excitation of
\methCN\ by He and \HH\ (Faure, private communication). Electron excitation 
is considered here as in \cite{Lis12Electrons}, using separate closed-form 
approximations for molecular ions and neutrals.

The excitation rate coefficients and our excitation calculations are 
not hyperfine-resolved and are just recalculations of the rotational 
partition function. Results of the excitation calculations are 
illustrated in Figure A.1 for hydrocarbons and CN-bearing species 
and in Figure A.2 for the oxygen-bearing species. The normalization
on the vertical axis is such that the integrated optical depth
of the transition in question corresponds to a total column density 
N $= 10^{11} \pcc$ (shown in each panel) but it is only the extent
of the variation across the horizontal axis that matters.  The default 
optical depth-column density conversion factor given in the 
next-to-last column of Table 2 corresponds to zero density
at the left and the maximum correction corresponds to the amount 
by which the curves have fallen at n(\HH) $ = 400 \pccc$.  The very 
lowest-lying transitions are quite sensitive to density variations 
while those lying higher may be nearly unaffected.  The excitation,
being dominated by electrons, is only weakly sensitive to the kinetic
temperature as shown in Figures A.1 and A.2 where the calculations
have been carried out for kinetic temperatures of 20, 40 and 60 K:
the different curves at these tempertures often overlap to the point 
that they are indistinguishable.

\begin{figure*}
\includegraphics[height=16cm]{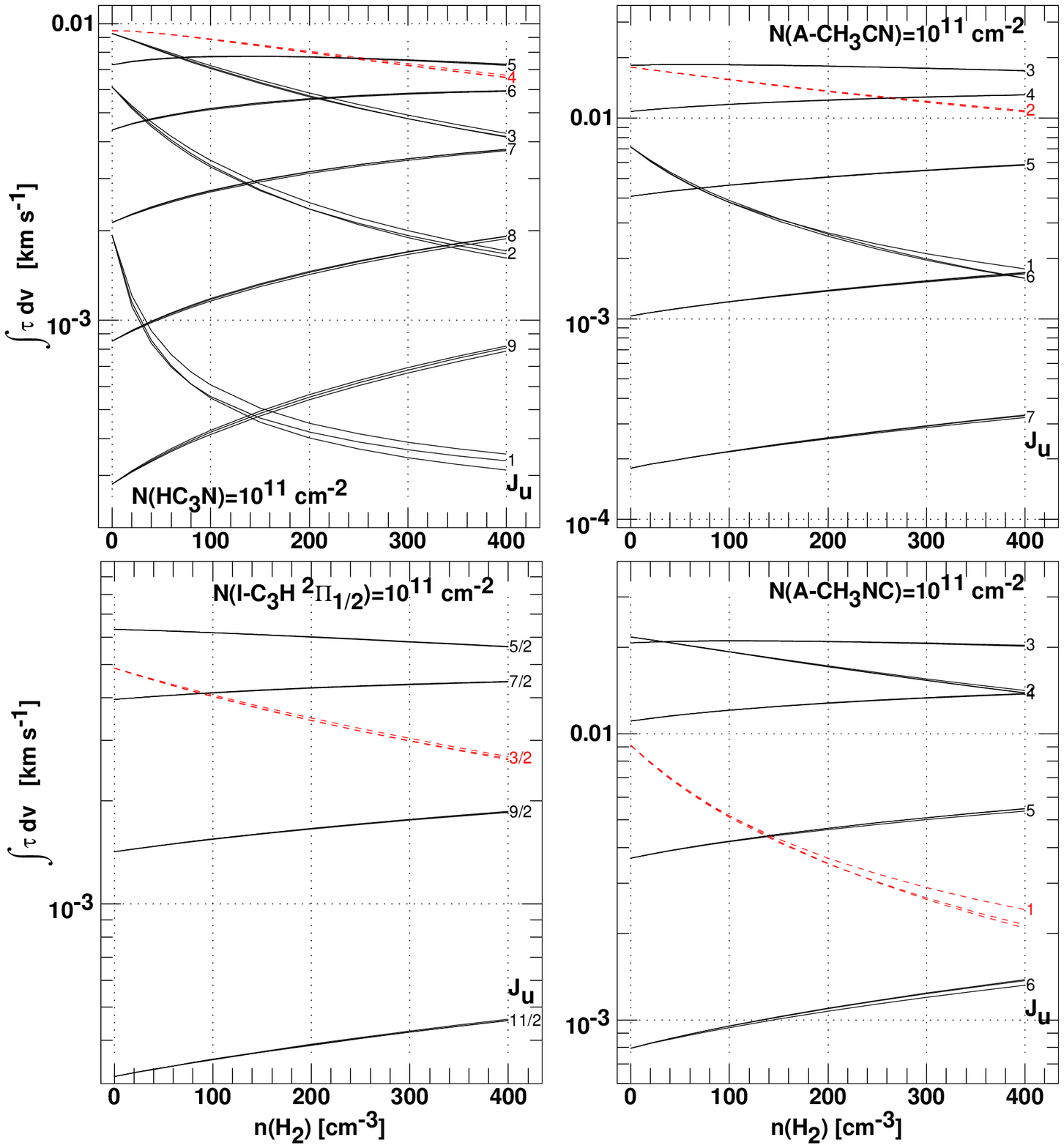}
  \caption{Integrated optical depth for rotational transitions of four
molecules observed in the course of this work, assuming a column density
of $10^{11}\pcc$ in each case.  The plots show the 
integrated optical depth of transitions whose upper-level quantum
number is shown at the right of each series of three curves. The three
curves for each transition correspond to calculations at 
kinetic temperatures of 20, 40 and 60 K and are often 
indistinguishable.  The excitation calculations include \HH\ and 
electrons for HC$_3$N and  He, \HH\ and electrons for \methCN, and 
only electrons otherwise, assuming an electron fraction 
n(e)/n(\HH) $= 3\times10^{-4}$.  The transition observed in this work 
is shown in red, dashed lines.}
\end{figure*}

\begin{figure*}
\includegraphics[height=8cm]{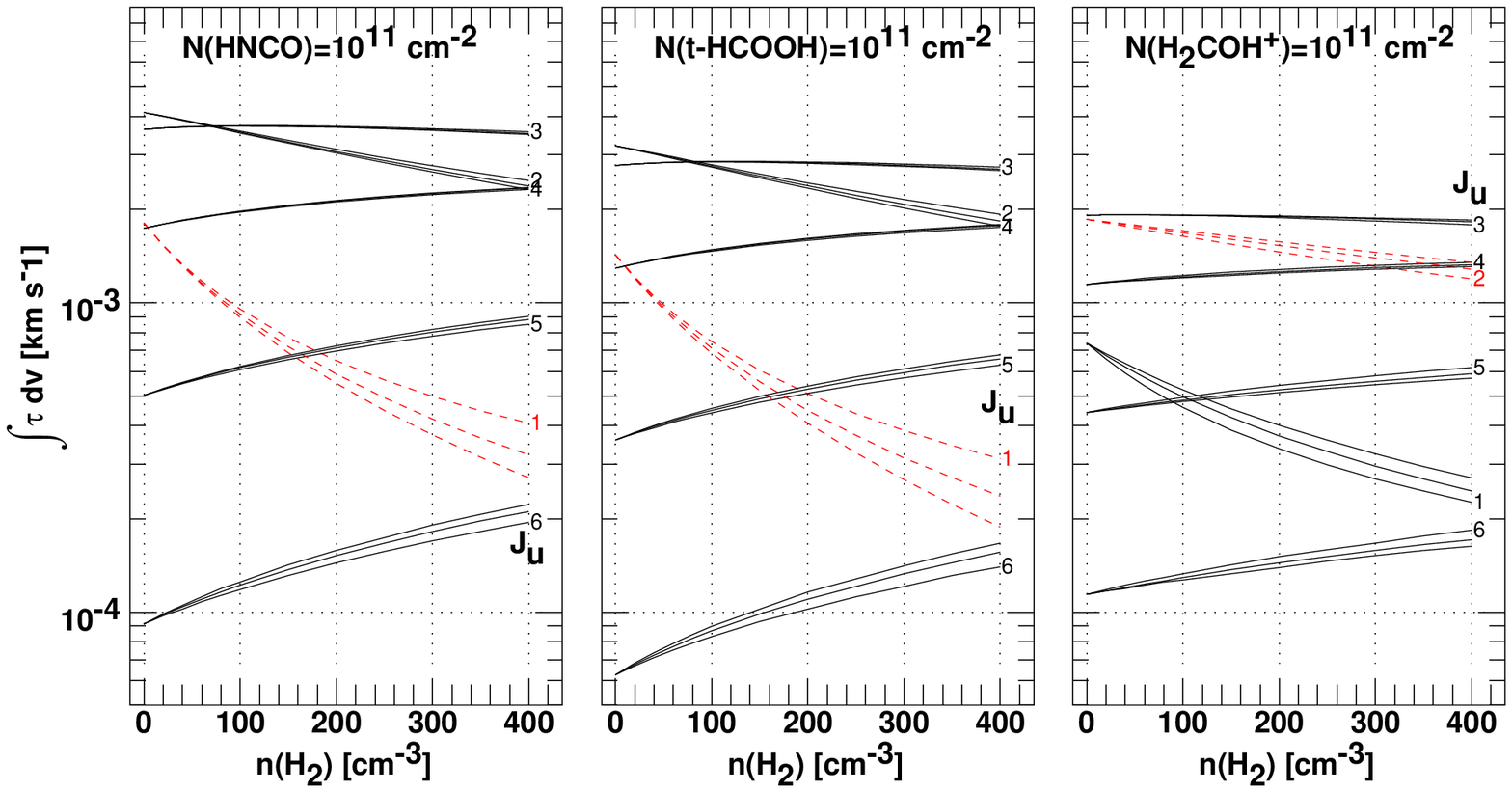}
  \caption{Integrated optical depth for rotational transitions of three
oxygen-bearing molecules observed in the course of this work, assuming a 
column density of $10^{11}\pcc$ in each case.  The plots show the 
integrated optical depth of transitions whose upper-level quantum
number is shown at the right of each series of three curves. The three
curves correspond to calculations at kinetic temperatures of 20, 40 and 60 K.
The calculations include electron excitation only, assuming
an electron fraction n(e)/n(\HH) $= 3\times10^{-4}$.  The transition 
observed in this work is shown in red, dashed lines.}
\end{figure*}

\bibliographystyle{aasjournal}

\end{document}